\documentclass[runningheads]{llncs}
\usepackage[T1]{fontenc}
\usepackage{graphicx}
\usepackage{multirow}
\usepackage{multicol}
\usepackage{url}
\usepackage{amssymb}
\usepackage{array}
\begin{document}
%
\title{Formula-Driven Data Augmentation and Partial Retinal Layer Copying for Retinal Layer Segmentation}
\titlerunning{Formula-Driven Data Augmentation and Partial Retinal Layer Copying}
%
%
%
\author{Tsubasa Konno\inst{1} \and
Takahiro Ninomiya\inst{2} \and
Kanta Miura\inst{1} \and
Koichi Ito\inst{1} \and
Noriko Himori\inst{2} \and
Parmanand Sharma\inst{2} \and
Toru Nakazawa\inst{2} \and
Takafumi Aoki\inst{1}
}
\authorrunning{T. Konno et al.}
\institute{Graduate School of Information Sciences, Tohoku University, 6-6-05, Aramaki Aza Aoba, Aoba-ku, Sendai-shi, Miyagi 9808579, Japan \\
\email{konno@aoki.ecei.tohoku.ac.jp} \and
Department of Ophthalmology, Graduate School of Medicine, Tohoku University, 1-1, Seiryomachi, Aoba-ku, Sendai-shi, Miyagi 9808574, Japan
}
\maketitle              
\begin{abstract}
  Major retinal layer segmentation methods from OCT images assume that the retina is flattened in advance, and thus cannot always deal with retinas that have changes in retinal structure due to ophthalmopathy and/or curvature due to myopia.
  To eliminate the use of flattening in retinal layer segmentation for practicality of such methods, we propose novel data augmentation methods for OCT images.
  Formula-driven data augmentation (FDDA) emulates a variety of retinal structures by vertically shifting each column of the OCT images according to a given mathematical formula.
  We also propose partial retinal layer copying (PRLC) that copies a part of the retinal layers and pastes it into a region outside the retinal layers.
  Through experiments using the OCT MS and Healthy Control dataset and the Duke Cyst DME dataset, we demonstrate that the use of FDDA and PRLC makes it possible to detect the boundaries of retinal layers without flattening even retinal layer segmentation methods that assume flattening of the retina.

\end{abstract}

\section{Introduction}
\label{sec:introduction}

Optical coherence tomography (OCT) is used in ophthalmologic examinations to non-invasively acquire the 3D structure of the retina.
The images acquired by OCT, i.e., OCT images, are used to measure the thickness of the retinal layers to diagnose ophthalmopathy that cause structural changes in the retina, such as glaucoma, age-related macular degeneration (AMD), and diabetic retinopathy.
The retina is composed of ten thin layers, and therefore it is difficult even for experts to manually measure the thickness of each layer.
Thus, retinal layer segmentation methods that detect each layer of the retina from OCT images have been investigated \cite{FCBRgraph2,FCBRgraph1,FCBRgraph3,FCBRgraph4,RelayNet,FCBR,FCBR2,SASR,SASR2}.
Graph-based methods \cite{FCBRgraph2,FCBRgraph1,FCBRgraph3,FCBRgraph4} have been primarily considered for retinal layer segmentation until around 2020, and now deep learning \cite{DL}-based methods \cite{RelayNet,FCBR,FCBR2,SASR,SASR2} have become the most common.
ReLayNet \cite{RelayNet} utilizes image segmentation based on U-Net \cite{UNet} to assign pixel-wise labels to retinal layers, background, and lesions.
Such methods similar to image segmentation produce errors such as isolated regions since the anatomical structure of the retinal layers cannot be considered.
The current major approaches in retinal layer segmentation detect the boundaries between retinal layers as straight lines from OCT images that are flattened to make the curved retina horizontal \cite{flatten}.
Fully convolutional boundary regression (FCBR) \cite{FCBR,FCBR2} detects the boundaries between retinal layers using 2D U-Net \cite{UNet}, while simultaneous alignment and surface regression (SASR) \cite{SASR,SASR2} detects the boundaries in three dimensions using the combined network of 2D encoder and 3D decoder.

\begin{figure}[t]
  \centering
  \includegraphics[width=.7\linewidth]{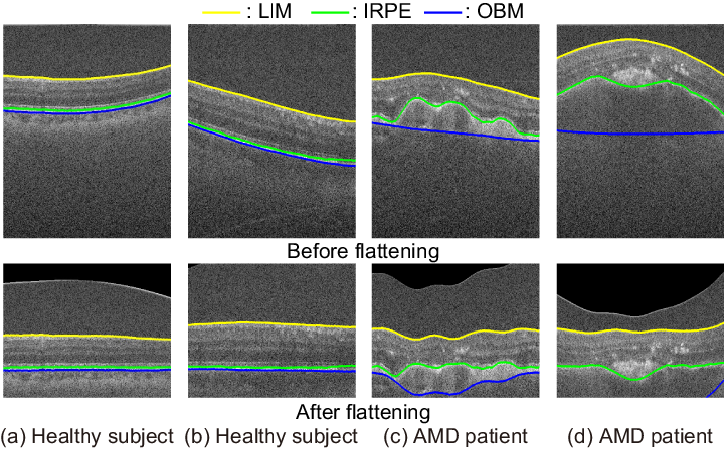}
  \caption{OCT images in the World's Largest Online Annotated SD-OCT dataset \cite{AMD} before and after flattening, where the yellow, green, and blue lines indicate the inner limiting membrane (ILM), the inner aspect of the retinal pigment epithelium drusen complex (IRPE), and the outer aspect of Bruch's membrane (OBM), respectively: (a) and (b) are the healthy subjects and (c) and (d) are the age-related macular degeneration (AMD) patients.}
  \label{fig:flattening}
\end{figure}

In flattening, the outer aspect of Bruch's membrane (OBM) is detected and the OCT image is deformed so that OBM is a horizontal straight line.
As shown in Fig. \ref{fig:flattening}, the images are correctly flattened in healthy subjects, while the images are not correctly flattened when the retinal structure is affected by ophthalmopathy.
Even in healthy subjects, the retina is significantly curved in the case of myopic eyes, and the pixel interpolation caused by the image deformation of flattening reduces the detection accuracy of the boundaries.
Hence, for practical situations, it is necessary to detect the boundaries of retinal layers without flattening.

For the above purpose, we need to train the network with a large number of training data with a variety of retinal structures.
To increase the variability of the training data, retinal layer segmentation methods use data augmentation such as horizontal flipping, vertical scaling, translation, noise, and contrast modulation \cite{SD-LayerNet,FCBR,SNetRNet,FCBR2,Li-BOE-2021,SASR,SASR2,RelayNet,Xie-OE-2022,Xie-BOE-2023}.
The networks used in retinal layer segmentation cannot learn a variety of retinal structures since these methods only apply typical variations to images.
In pathological retinal region segmentation and chorio-retinal segmentation, several methods have been proposed to improve the accuracy using the generative adversarial network (GAN) \cite{GAN1,GAN2,GAN3}.
Generating OCT images with a variety of retinal structures using GAN can increase the variability of the training data.
On the other hand, it is necessary to annotate the generated images and GAN does not always generate high-quality OCT images.

In this paper, we propose formula-driven data augmentation (FDDA), which is a data augmentation method that emulates a variety of retinal structures based on mathematical formulas for training data.
FDDA changes the retinal structure by vertically shifting each column of the OCT images according to a given mathematical formula.
For example, zero-order FDDA emulates the position of the retina, first-order FDDA emulates the tilt of the retina, and second-order FDDA emulates the curvature of the retina.
Since FDDA is based on a mathematical shift, the labels after data augmentation can be obtained by shifting the boundary labels associated with the training data in the same way.
In addition to FDDA, we also propose partial retinal layer copying (PRLC), which is a data augmentation method that copies a part of the retinal layers and pastes it into a region outside the retinal layers.
Since structures similar to the retinal layers may be falsely detected, PRLC allows us to take into account the global structure of the retinal layers by providing partial fake retinal layers.
Through experiments using the OCT MS and Healthy Control (MSHC) dataset\footnote[1]{\url{https://iacl.ece.jhu.edu/index.php/Resources}} \cite{MS} and the Duke Cyst DME (Duke DME) dataset\footnote[2]{\url{https://people.duke.edu/~sf59/Chiu_BOE_2014_dataset.htm}} \cite{FCBRgraph2}, we demonstrate that the use of the proposed data augmentation makes it possible to detect the boundaries of retinal layers without flattening even retinal layer segmentation methods that assume flattening of the retina.

\section{Method}
\label{sec:method}

This section describes the details of FDDA and PRLC proposed in this paper.

\subsection{Formula-Driven Data Augmentation (FDDA)}
\label{sec:fdda}

\begin{figure}[t]
  \centering
  \includegraphics[width=.85\linewidth]{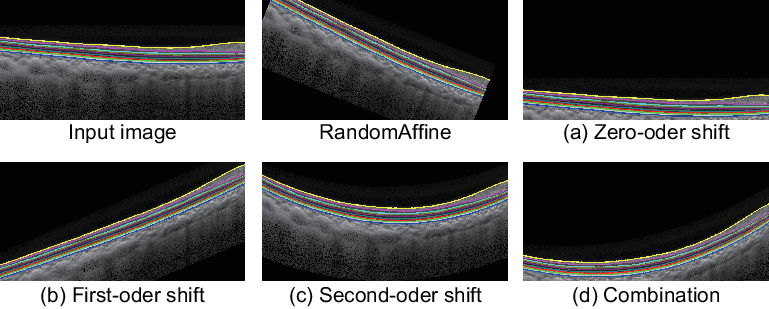}
  \caption{Examples of applying FDDA to OCT images in MSHC, where each of the zero-order, first-order, and second-order shifts is applied to the input image for simplicity: (a) the zero-order shift $\Delta_0(n_2)$, (b) the first-order shift $\Delta_1(n_2)$, (c) the second-order shift $\Delta_2(n_2)$, and (d) the combined shift $\Delta(n_2)$, and an example of applying RandomAffine for comparison.
  Colored lines on each image indicate the annotated boundaries between the retinal layers.}
  \label{fig:fdda_image}
\end{figure}

FDDA is a data augmentation method that shifts each column of an OCT image according to a simple combination of $N$th-order functions.
In FDDA, the image center is the origin of $N$th-order functions, and the amount of pixel shift is determined based on these functions.
Since this process only shifts the pixels by the amount of shift determined by simple functions, the labels of the OCT image after data augmentation can be easily obtained by shifting the boundary labels in the same way.
Let $I(n_1,n_2)$ be a 2D OCT image, where $1\leq n_1 \leq N_1$ and $1\leq n_2 \leq N_2$.
The upper left corner of the image is $(n_1,n_2)=(1,1)$, and the vertical and horizontal directions of the image are the $n_1$ and $n_2$ axes, respectively.
Let $(c_1,c_2) = (\lfloor N_1/2\rfloor+1,\lfloor N_2/2\rfloor+1)$ be the center pixel of the image, where $\lfloor\cdot\rfloor$ denotes the floor function.
Let $a(k)$ be a parameter for the $k$th-order shift, then the zero-order shift $\Delta_0(n_2)$, the first-order shift $\Delta_1(n_2)$, and the second-order shift $\Delta_2(n_2)$ for the $n_1$ axis are defined by
\begin{eqnarray}
  \Delta_0(n_2) &=& a(0),\\
  \Delta_1(n_2) &=& a(1)(n_2-c_2),\\
  \Delta_2(n_2) &=& a(2)(n_2-c_2)^2,
\end{eqnarray}
respectively.
As a result, the combined shift $\Delta(n_2)$ from zero-order to $N$th-order on the $n_1$ axis is defined by
\begin{eqnarray}
  \Delta(n_2) = \sum_{k=0}^N\Delta_k(n_2) = \sum_{k=0}^{N}a(k)(n_2-c_2)^k.
\end{eqnarray}
That is, the amount of shift $\Delta(n_2)$ in FDDA is defined by a linear basis expansion with $N$th-order equations as its basis functions.
The image $I'(n_1,n_2)$ obtained by FDDA is given by
\begin{equation}
  \label{eq:fdda}
  I'(n_1,n_2) =
  \left\{
  \begin{array}{cc}
    I(n_1+\lfloor\Delta(n_2)+0.5\rfloor,n_2) & {\rm if} \ 1\leq n_1+\lfloor\Delta(n_2)+0.5\rfloor \leq N_1\\
    0 & {\rm otherwise}
  \end{array}
  \right..
\end{equation}
Since background regions in OCT images have pixel values close to zero, we apply zero-padding to pixels that are not assigned pixel values by shifting, as in Eq. (\ref{eq:fdda}).
FDDA can emulate retinal structures of various shapes by changing a set of parameters $a(k)$ $(k=0\sim N)$.
Note that FDDA can be easily extended to higher orders, although this paper is limited to the second order, i.e., $N=2$.
When applying FDDA to volume data, the same shift is applied in each slice image.
Fig. \ref{fig:fdda_image} shows examples of applying FDDA to OCT images in MSHC \cite{MS}, where we use each of the zero-order, first-order, and second-order shifts for simplicity.
By applying FDDA to flat retinal OCT images of a healthy subject, we can obtain OCT images of a variety of retinal structures.
For comparison, we also show an example of applying RandomAffine, which is a data augmentation method similar to FDDA, to an OCT image.
The retina is deformed unnaturally compared to the images with FDDA applied.

\begin{figure}[t]
  \centering
  \includegraphics[width=.8\linewidth]{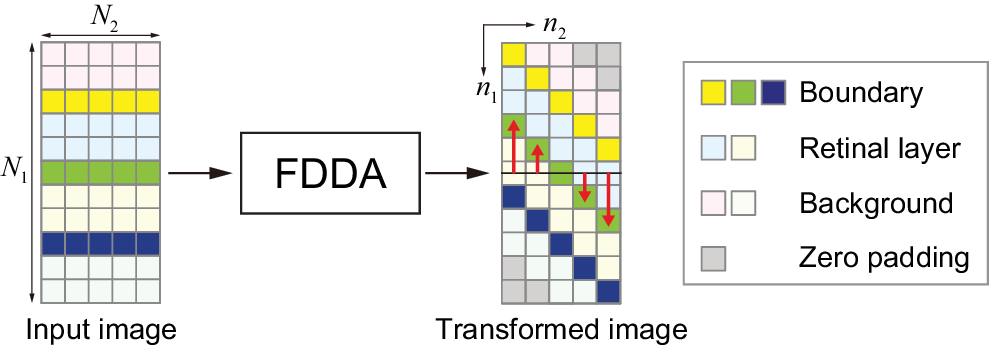}
  \caption{Overview of FDDA using only the first-order shift $\Delta_1(n_2)$ with $a(1)=1$.}
  \label{fig:fdda_flow}
\end{figure}

\subsection{Partial Retinal Layer Copying (PRLC)}
\label{sec:rlc}

\begin{figure}[t]
  \centering
  \includegraphics[width=.85\linewidth]{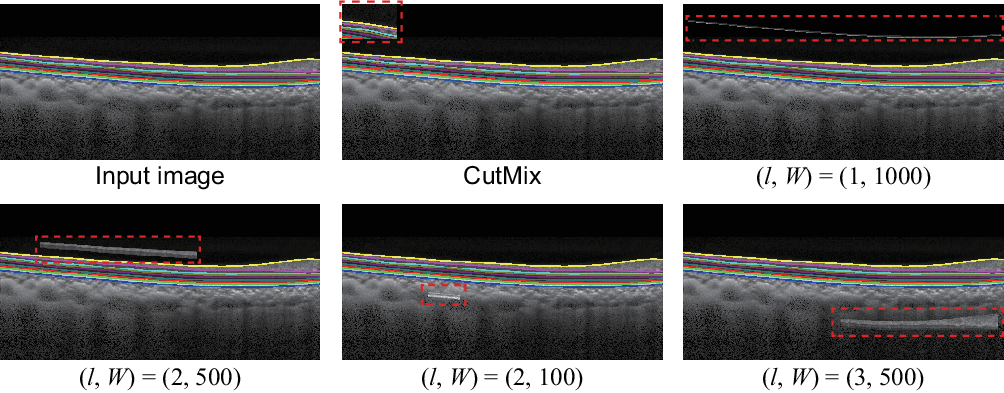}
  \caption{Examples of applying PRLC to OCT images in MSHC, where the red dashed box indicates the pasted retinal layer area.
  An example of applying CutMix is also shown for comparison.
  Colored lines on each image indicate the annotated boundaries between the retinal layers.}
  \label{fig:rlc_image}
\end{figure}

PRLC is a data augmentation method that copies a part of the retinal layers and pastes it into a region outside the retinal layers.
The parameters in PRLC are the number of retinal layers $l$ to be copied and the width $W$ of the retinal layers.
In this paper, $l$ is set to 1 to 3 and $W$ is set to 20 to $N_2$, which are selected at random from the corresponding ranges.
First, the target retinal layer is randomly determined and the adjacent retinal layers are selected according to $l$.
Next, the retinal layer region to be copied is determined according to $W$.
Then, we paste the above retinal layer region at a random position in the background region where no retinal layer labels are assigned.
If there is no space in the background region to paste the retinal layer region, we repeat the process from the first step.
When applying PRLC to volume data, the parameters and paste position are the same for each slice image, and the retinal layer region copied from each slice image is pasted.
There are several data augmentation methods similar to PRLC, such as Cutout \cite{cutout}, RandomErasing \cite{randerase}, and CutMix \cite{cutmix}.
Cutout and RandomErasing reduce overfitting and train in consideration of occlusion by masking a part of the image.
CutMix is an extended version of Cutout and RandomErasing, which mixes two images and assigns a probability to each label according to its area.
On the other hand, PRLC reduces false detection in the background area by copying the retinal layers to the background area and keeping the label assignments unchanged.
Fig. \ref{fig:rlc_image} shows examples of applying PRLC to OCT images and an example of applying CutMix for comparison in MSHC \cite{MS}.
The network is trained to avoid false detection in the background region since the retinal layer region pasted by PRLC becomes noise.

\section{Experiments}
\label{sec:experiments}

In this section, we demonstrate the effectiveness of the proposed method for retinal layer segmentation by applying FDDA and PRLC to FCBR \cite{FCBR,FCBR2} and SASR \cite{SASR,SASR2}.
In the following, we describe the dataset used in the experiments, the experimental conditions, and the experimental results.

\subsection{Datasets}
\label{sec:datasets}

We use the MSHC dataset\footnotemark[1] \cite{MS} and the Duke DME dataset\footnotemark[2] \cite{FCBRgraph2} in the experiments.
MSHC consists of OCT images acquired from 14 healthy subjects and 21 MS patients.
Each OCT image consists of 49 B-scan images with $496\times 1,024$ pixels.
This dataset provides 9 boundary labels as the ground truth.
We use 6 healthy subjects and 9 MS patients from the end of the subject number for training and validation, and the remaining for test, as in FCBR \cite{FCBR}.
Of the 15 subjects in the training and validation data, one healthy subject and two MS patients are used for validation in ascending order of subject number.
Duke DME consists of OCT images acquired from 10 DME patients.
Each OCT image consists of 61 B-scan images with $496\times 768$ pixels.
This dataset provides 8 boundary labels and an edema mask as the ground truth.
Note that 61 images are provided for each patient, of which 11 images with boundary labels are used in the experiment.
Due to the small number of data in this dataset, we conduct 5-fold cross-validation, where 6 patients for training, 2 patients for validation, and 2 patients for test.

\subsection{Experimental Condition}
\label{sec:condition}

In this experiment, we evaluate the accuracy of boundary detection by introducing FDDA and PRLC to FCBR \cite{FCBR,FCBR2} and SASR \cite{SASR,SASR2} without flattening the retina.
FCBR \cite{FCBR,FCBR2} detects the boundaries between retinal layers using 2D U-Net.
Since the code for FCBR is not publicly available, our reproduced code is used in the experiments.
SASR \cite{SASR,SASR2} detects the boundaries between retinal layers in three dimensions using the combined network of 2D encoder and 3D decoder.
We use the publicly available code of SASR\footnote[3]{\url{https://github.com/ccarliu/Retinal-OCT-LayerSeg/tree/following-work}}.
FCBR and SASR assume that OCT images are flattened in preprocessing.
Since flattening\footnote[4]{\url{https://github.com/YufanHe/oct_preprocess}} is not always applicable, we demonstrate that FDDA and PRLC improve the practicality of boundary detection.
Image normalization is performed in the same way as the conventional methods \cite{FCBR,FCBR2,SASR,SASR2}, while no preprocessing is performed for SASR on Duke DME.
FCBR \cite{FCBR,FCBR2} employs horizontal flipping and vertical scaling, while SASR \cite{SASR,SASR2} uses horizontal flipping only on Duke DME.
When FDDA and/or PRLC are introduced, they are used in combination with these data augmentation methods.
FDDA and PRLC are applied with a probability of 50\% each.
In FDDA, $N=2$, $a(0)$ is the range of retinal layers contained within each image, $-0.5 \leq a(1) \leq 0.5$ for both datasets, $-0.0002 \leq a(2) \leq 0.0002$ for MSHC, and $-0.00068 \leq a(2) \leq 0.00068$ for Duke DME. 
We demonstrate the effectiveness of FDDA and PRLC by comparing them to RandomAffine and CutMix as data augmentation.
The accuracy of each method is evaluated by the mean absolute distance (MAD) between the detected boundary and the ground truth.
MAD is calculated as the product of the mean error [pixels] in each column and the image resolution [$\mu$m/pixel] in the column direction, where MSHC is 3.9 $\mu$m/pixel and Duke DME is 3.87 $\mu$m/pixel in the experiments.
The smaller the value of MAD, the higher the accuracy of the boundary detection.
The standard deviation (SD) of MAD per subject is also evaluated.

\subsection{Experimental Results and Discussion}
\label{sec:results}

\begin{figure}[t]
  \centering
  \includegraphics[width=.9\linewidth]{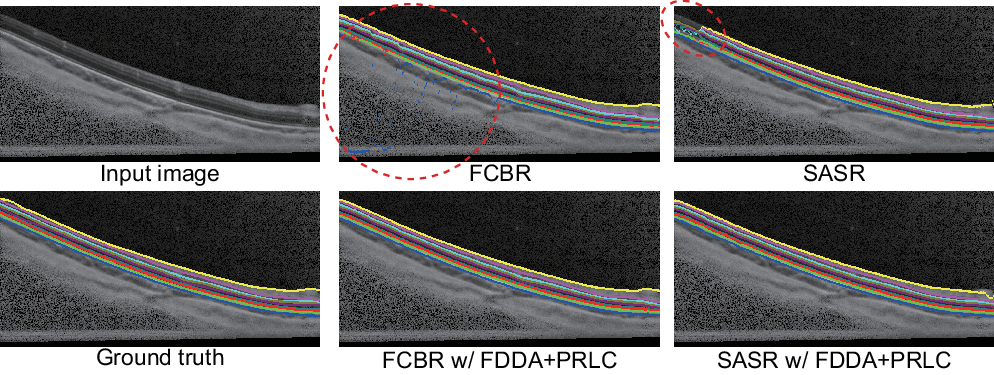}
  \caption{Detection results of the retinal layer boundaries from the OCT images of MSHC using each method.
  The red dotted circles indicate the region with large errors.}
  \label{fig:result_mshc}
\end{figure}

Table \ref{tb:results} shows the quantitative evaluation results of the boundary detection for each method.
For reference, we include the results of FCBR \cite{FCBR} and SASR \cite{SASR2} with flattening from the corresponding literature.
Note that the experimental conditions for both are different for Duke DME.
Both FCBR and SASR significantly degrade the detection accuracy without flattening.
Introducing RandomAffine or CutMix does not improve detection accuracy.
On the other hand, introducing FDDA or PRLC improves the detection accuracy of both methods.
For MSHC, introducing FDDA and PRLC improves the detection accuracy comparable when applying flattening.
For Duke DME, introducing FDDA and PRLC improves the detection accuracy higher than when applying flattening.
Fig. \ref{fig:result_mshc} shows the results of detecting the retinal layer boundaries from the OCT images of MSHC using each method.
FCBR \cite{FCBR,FCBR2} makes significant errors in detecting the blue boundary in the region surrounded by the red dotted circle.
SASR \cite{SASR,SASR2} makes significant errors at the left side of the image in the regions surrounded by the red dotted circles.
The reasons for these errors may be that these methods do not deal with large changes in retinal structure, and that background noise is falsely detected as boundaries.
Introducing FDDA and PRLC prevents such false detection, and both methods can detect boundaries that are close to the ground truth.
As a result, the use of FDDA and PRLC makes it possible to detect the boundaries without flattening the retina with the same or higher accuracy than when flattening is performed.

\begin{table}[t]
  \caption{Experimental results of each method for MSHC and Duke DME.
  The units for MAD and SD are $\mu$m.}
  \label{tb:results}
  \centering
  \begin{tabular}{lccccccc}
    \hline
    Method & Flattening & RandomAffine & CutMix & FDDA & PRLC & MSHC & Duke DME\\
    \hline
    FCBR \cite{FCBR}      & \checkmark & & & & & 2.83$\pm$0.99 & 6.70\\
    \hline
    \multirow{7}{*}{FCBR} & \checkmark & & & & & 2.92$\pm$0.55 & 6.59$\pm$2.43 \\
                          & & & & & &            3.87$\pm$4.28 & 6.94$\pm$2.97 \\
                          & & \checkmark & & & & 3.76$\pm$0.68 & 6.44$\pm$3.14 \\
                          & & & \checkmark & & & 3.52$\pm$0.61 & 6.68$\pm$2.85 \\
                          & & & & \checkmark & & 2.92$\pm$0.57 & 6.04$\pm$2.58 \\
                          & & & & & \checkmark & 3.16$\pm$1.13 & 6.32$\pm$2.54 \\
                          & & & & \checkmark & \checkmark & {\bf 2.84$\pm$0.50} & {\bf 5.97$\pm$2.59}\\
    \hline
    SASR \cite{SASR2}     & \checkmark & & & & & 2.77$\pm$0.51 & 6.37$\pm$1.01\\
    \hline
    \multirow{5}{*}{SASR} & \checkmark & & & & & 2.87$\pm$0.51 & 6.54$\pm$2.96 \\
                          & & & & & &            3.05$\pm$0.87 & 6.34$\pm$1.67\\
                          & & & & \checkmark & & 2.92$\pm$0.61 & 5.84$\pm$1.33\\
                          & & & & & \checkmark & 2.99$\pm$0.72 & 6.10$\pm$1.40\\
                          & & & & \checkmark & \checkmark & {\bf 2.90$\pm$0.58} & {\bf 5.83$\pm$1.37}\\
    \hline
  \end{tabular}
\end{table}

\section{Conclusion}
\label{sec:conclusion}

In this paper, we proposed novel data augmentation methods: FDDA that emulates a variety of retinal structures based on mathematical formulas and PRLC that copies a part of the retinal layers and pastes it into a region outside the retinal layers.
Through experiments using public datasets, we demonstrated that the use of the proposed methods makes it possible not to apply flattening to the OCT image in retinal layer segmentation methods.
\begin{credits}
\subsubsection{\discintname}
The authors have no competing interests to declare that are relevant to the content of this article.
\end{credits}
%
%
%
\bibliographystyle{splncs04}
\bibliography{Paper-0031}

\begin{thebibliography}{10}
\providecommand{\url}[1]{\texttt{#1}}
\providecommand{\urlprefix}{URL }
\providecommand{\doi}[1]{https://doi.org/#1}

\bibitem{FCBRgraph2}
Chiu, S.J., Allingham, M.A., Mettu, P.S., Cousins, S.W., Izatt, J.A., Farsiu,
  S.: Kernel regression based segmentation of optical coherence tomography
  images with diabetic macular edema. Biomed. Opt. Express  \textbf{6}(4),
  1172--1194 (Apr 2015)

\bibitem{FCBRgraph1}
Chiu, S.J., Li, X.T., Nicholas, P., Toth, C.A., Izatt, J.A., Farsiu, S.:
  Automatic segmentation of seven retinal layers in {SDOCT} images congruent
  with expert manual segmentation. Opt. Express  \textbf{18}(18),  19413--19428
  (Aug 2010)

\bibitem{cutout}
Devries, T., Taylor, G.W.: Improved regularization of convolutional neural
  networks with {Cutout}. CoRR  \textbf{abs/1708.04552}, ~1--8 (Aug 2017)

\bibitem{AMD}
Farsiu, S., Chiu, S.J., O'Connell, R.V., Folgar, F.A., Yuan, E., Izatt, J.A.,
  Toth, C.A.: Quantitative classification of eyes with and without intermediate
  age-related macular degeneration using optical coherence tomography.
  Ophthalmology  \textbf{121}(1),  162--172 (Jan 2014)

\bibitem{SD-LayerNet}
Fazekas, B., Aresta, G., Lachinov, D., Riedl, S., Mai, J., Schmidt-Erfurth, U.,
  Bogunović, H.: {SD}-{L}ayer{N}et: {S}emi-supervised retinal layer
  segmentation in {OCT} using disentangled representation with anatomical
  priors. Proc. Int'l Conf. Medical Image Computing and Computer Assisted
  Intervention pp. 320--329 (Sep 2022)

\bibitem{DL}
Goodfellow, I., Bengio, Y., Courville, A.: Deep Learning. MIT Press (2016)

\bibitem{FCBR}
He, Y., Carass, A., Jedynak, B.M., Solomon, S.D., Saidha, S., Calabresi, P.A.,
  Prince, J.L.: Fully convolutional boundary regression for retina {OCT}
  segmentation. Proc. Int'l Conf. Medical Image Computing and Computer Assisted
  Intervention pp. 120--128 (Oct 2019)

\bibitem{SNetRNet}
He, Y., Carass, A., Liu, Y., Jedynak, B.M., Solomon, S.D., Saidha, S.,
  Calabresi, P.A., Prince, J.L.: Deep learning based topology guaranteed
  surface and {MME} segmentation of multiple sclerosis subjects from retinal
  {OCT}. Biomed. Opt.Express  \textbf{10}(10),  5042--5058 (Oct 2019)

\bibitem{FCBR2}
He, Y., Carass, A., Liu, Y., Jedynak, B.M., Solomon, S.D., Saidha, S.,
  Calabresi, P.A., Prince, J.L.: Structured layer surface segmentation for
  retina {OCT} using fully convolutional regression networks. Medical Image
  Analysis  \textbf{68},  101856 (Feb 2021)

\bibitem{MS}
He, Y., Carass, A., Solomon, S.D., Saidha, S., Calabresi, P.A., Prince, J.L.:
  Retinal layer parcellation of optical coherence tomography images: {D}ata
  resource for multiple sclerosis and healthy controls. Data Brief
  \textbf{22},  601--604 (Feb 2018)

\bibitem{FCBRgraph3}
Karri, S.P.K., Chakraborthi, D., Chatterjee, J.: Learning layer-specific edges
  for segmenting retinal layers with large deformations. Biomed. Opt. Express
  \textbf{7}(7),  2888--2901 (Jul 2016)

\bibitem{GAN2}
Kugelman, J., Alonso-Caneiro, D., Read, S.A., Vincent, S.J., Chen, F.K.,
  Collins, M.J.: Data augmentation for patch-based {OCT} chorio-retinal
  segmentation using generative adversarial networks. Neural Computing and
  Applications  \textbf{33}(13),  7393--7408 (Mar 2021)

\bibitem{GAN3}
Kugelman, J., Alonso-Caneiro, D., Read, S.A., Vincent, S.J., Collins, M.J.:
  Enhanced {OCT} chorio-retinal segmentation in low-data settings with
  semi-supervised {GAN} augmentation using cross-localisation. Computer Vision
  and Image Understanding  \textbf{237}(103852),  1--14 (Dec 2023)

\bibitem{flatten}
Lang, A., Carass, A., Hauser, M., Sotirchos, E.S., Calabresi, P.A., Ying, H.S.,
  Prince, J.L.: Retinal layer segmentation of macular {OCT} images using
  boundary classification. Biomed. Opt. Express  \textbf{4}(7),  1133--1152
  (Jul 2013)

\bibitem{Li-BOE-2021}
Li, J., Jin, P., Zhu, J., Zou, H., Xu, X., Tang, M., Zhou, M., Gan, Y., He, J.,
  Ling, Y., Su, Y.: Multi-scale {GCN}-assisted two-stage network for joint
  segmentation of retinal layers and discs in peripapillary {OCT} images.
  Biomed. Opt. Express  \textbf{12}(4),  2204--2220 (Apr 2021)

\bibitem{SASR}
Liu, H., Wei, D., Lu, D., Li, Y., Ma, K., Wang, L., Zheng, Y.: Simultaneous
  alignment and surface regression using hybrid {2D-3D} networks for {3D}
  coherent layer segmentation of retina {OCT} images. Proc. Int'l Conf. Medical
  Image Computing and Computer Assisted Intervention pp. 108--118 (Sep 2021)

\bibitem{SASR2}
Liu, H., Wei, D., Lu, D., Tang, X., Wang, L., Zheng, Y.: Simultaneous alignment
  and surface regression using hybrid 2{D}–3{D} networks for 3{D} coherent
  layer segmentation of retinal {OCT} images with full and sparse annotations.
  Medical Image Analysis  \textbf{91}(103019),  1--14 (Jan 2024)

\bibitem{GAN1}
Mahapatra, D., Bozorgtabar, B., Shao, L.: Pathological retinal region
  segmentation from {OCT} images using geometric relation based augmentation.
  Proc. IEEE/CVF Conf. Computer Vision and Pattern Recognition pp. 9611--9620
  (Jun 2020)

\bibitem{FCBRgraph4}
Rathke, F., Desana, M., Schnörr, C.: Locally adaptive probabilistic models for
  global segmentation of pathological {OCT} scans. Proc. Int'l Conf. Medical
  Image Computing and Computer Assisted Intervention  \textbf{10433},  177--184
  (Sep 2017)

\bibitem{UNet}
Ronneberger, O., Fischer, P., Brox, T.: {U-Net}: {C}onvolutional networks for
  biomedical image segmentation. Proc. Int'l Conf. Medical Image Computing and
  Computer Assisted Intervention pp. 234--241 (Oct 2015)

\bibitem{RelayNet}
Roy, A., Conjeti, S., Karri, S., Sheet, D., Katouzian, A., Wachinger, C.,
  Navab, N.: {ReLayNet}: {R}etinal layer and fluid segmentation of macular
  optical coherence tomography using fully convolutional networks. Biomed. Opt.
  Express  \textbf{8}(8),  3627--3642 (Aug 2017)

\bibitem{Xie-OE-2022}
Xie, H., Pan, Z., Zhou, L., Zaman, F.A., Chen, D.Z., Jonas, J.B., Xu, W., Wang,
  Y.X., Wu, X.: Globally optimal {OCT} surface segmentation using a constrained
  {IPM} optimization. Opt. Express  \textbf{30}(2),  2453--2471 (Jan 2022)

\bibitem{Xie-BOE-2023}
Xie, H., Xu, W., Wang, Y.X., Wu, X.: Deep learning network with differentiable
  dynamic programming for retina {OCT} surface segmentation. Biomed. Opt.
  Express  \textbf{14}(7),  3190--3202 (Jul 2023)

\bibitem{cutmix}
Yun, S., Han, D., Oh, S.J., Chun, S., Choe, J., Yoo, Y.: {CutMix}:
  {R}egularization strategy to train strong classifiers with localizable
  features. Proc. IEEE/CVF Int'l Conf. Computer Vision pp. 6023--6032 (Oct
  2019)

\bibitem{randerase}
Zhong, Z., Zheng, L., Kang, G., Li, S., Yang, Y.: Random erasing data
  augmentation. Proc. AAAI Conf. Artificial Intelligence  \textbf{34}(7),
  13001--13008 (Feb 2020)

\end{thebibliography}

\end{document}